\documentclass[sigplan, nonacm]{acmart}

\usepackage{graphicx}
\usepackage{subcaption}

\usepackage{multirow}
\usepackage{amsmath}
\usepackage{amsmath,amsfonts}
\usepackage[ruled,vlined,linesnumbered]{algorithm2e}
\usepackage[normalem]{ulem}
\SetKwInput{KwIn}{Input}      
\SetKwInput{KwOut}{Output}

\SetKwFunction{US}{UniSample}
\SetKwFunction{Fn}{Function}

\begin{document}

\title{Tangram: Accelerating Serverless LLM Loading through GPU Memory Reuse and Affinity}

\author{Wenbin Zhu}
\affiliation{%
  \institution{Shandong University}
  \country{China}}
\email{wenbinzhu@mail.sdu.edu.cn}

\author{Zhaoyan Shen}
\affiliation{%
  \institution{Shandong University}
  \country{China}}
\email{shenzhaoyan@sdu.edu.cn}

\author{Zili Shao}
\affiliation{%
  \institution{The Chinese University of Hong Kong}
  \country{China}}
\email{shao@cse.cuhk.edu.hk}

\author{Hongjun Dai}
\affiliation{%
  \institution{Shandong University}
  \country{China}}
\email{dahogn@sdu.edu.cn}

\author{Feng Chen}
\affiliation{%
  \institution{Indiana University Bloomington}
  \country{USA}}
\email{fchen25@iu.edu}

\begin{abstract}
Serverless Large Language Models (LLMs) have emerged as a cost-effective solution for deploying AI services by enabling a ``pay-as-you-go'' pricing model through GPU resource sharing. However, cold-start latency, especially the model loading phase, has become a critical performance bottleneck, as it scales linearly with model size and severely limits the practical deployment of large-scale LLM services. 
This paper presents {\it Tangram}, a novel system that accelerates Serverless LLM loading through efficient GPU memory reuse. By leveraging the unused GPU memory to retain model parameters, Tangram significantly reduces model transfer time and cold-start latency. Its design includes three key components: unified GPU memory pool for tensor-level parameter sharing across models, on-demand KV cache allocation for dynamic memory management, and GPU-affinity-aware scheduling for maximizing resource utilization. These techniques collectively address the critical challenges of inefficient memory usage and the cold-start problem in Serverless LLM platforms. 
We have implemented a fully functional prototype, and experiments show that Tangram achieves up to $6.2\times$ faster loading and reduces Time-To-First-Token (TTFT) during cold-start by 23–55\% over state-of-the-art methods.

\end{abstract}

\maketitle

\section{Introduction}
\label{sec:intro}
\begin{figure}
    \centering
    \includegraphics[width=0.5\textwidth]{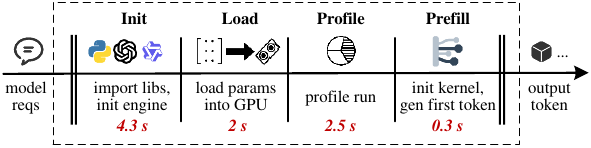}
    \caption{Multi-phase initialization of Serverless LLM and Time-To-First-Token (TTFT) composition for GPT-20B.}
    \label{fig:startup}
\end{figure}

\begin{figure}
    \centering
    \begin{subfigure}[b]{0.95\linewidth}
        \centering
        \includegraphics[width=\linewidth]{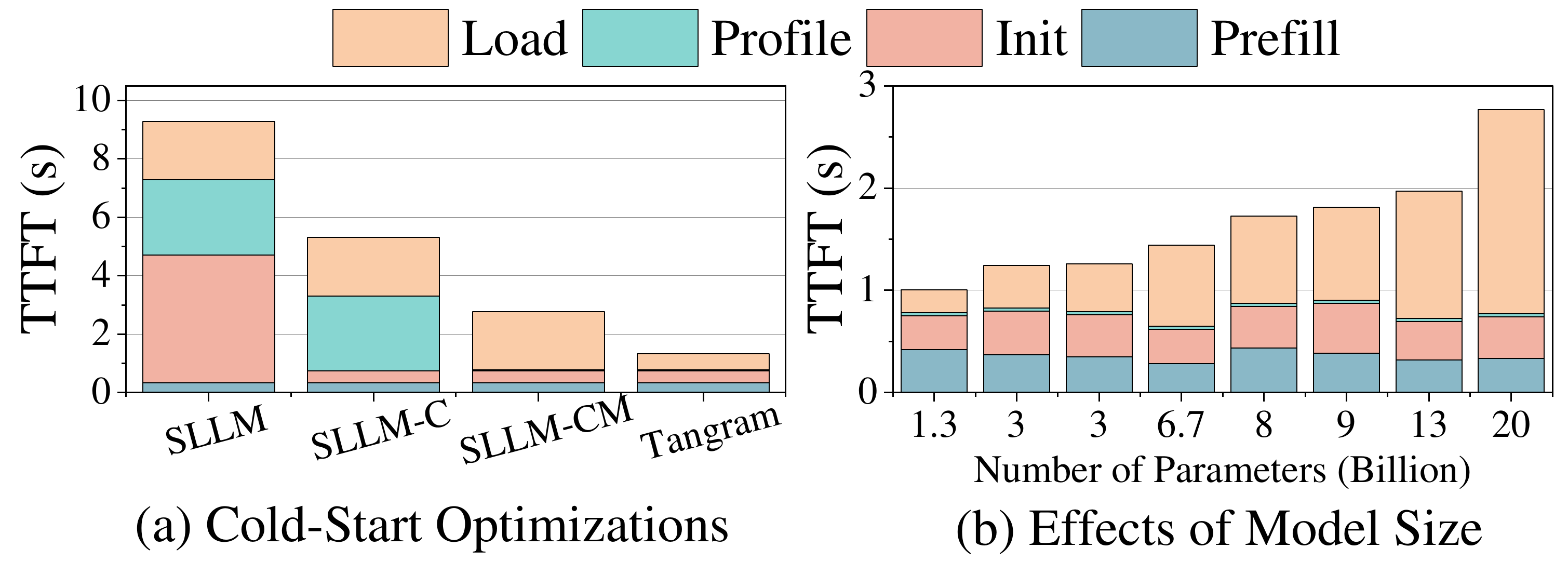}
    \end{subfigure}
    \caption{(a) TTFT breakdown for GPT-20B under different cold-start optimizations. (b) TTFT breakdown of SLLM-CM across models of varying parameter sizes.}
    \label{fig:cold-start-optimizations}
\end{figure}

In recent years, the demand for Large Language Model (LLM) services has rapidly surged across nearly all sectors. A critical challenge for many users and organizations is the prohibitively high deployment costs.
A promising emerging solution is {\it Serverless LLM}~\cite{fu2024serverlessllm}, represented by Google Vertex AI~\cite{GoogleVertexAI}, Amazon Bedrock~\cite{AmazonDebrok}, Alibaba PAI-EAS~\cite{AliPAI}, and Hyperbolic~\cite{hyperbolic}. 
As an online service platform, Serverless LLM provides on-demand services, allowing users to share GPU resources and deploy their LLM models. By multiplexing the expensive hardware infrastructure, the system dynamically switches between LLM models and reallocates GPU resources among users, enabling a flexible, efficient ``pay-as-you-go'' pricing model that significantly lowers the operation costs for users~\cite{serverlessinference, fu2024serverlessllm, serverlessml-coldstart-transform}.

However, due to the ever-growing parameter size and the complexity of multi-phase initialization, an increasingly prominent challenge is that Serverless LLMs tend to suffer from significant model-switching latency, referred to as the {\it cold-start problem}, which severely impacts latency-sensitive applications~\cite{fu2024serverlessllm,medusa,tidal,yu2025prism, serverless-coldstart-share, serverless-coldstart-init}. 

In a typical Serverless LLM deployment, initiating a new model instance involves four phases: 
\emph{Init}, which configures the runtime environment, including Python libraries and inference engines; 
\emph{Load}, which transfers the model parameters into GPU memory; 
\emph{Profile}, which performs a profiling run to determine the maximum available KV cache size; and 
\emph{Prefill}, which initializes the CUDA kernels and executes the initial forward pass. 
After these steps, the activated model generates its first inference token. Consequently, the {\it Time-to-First-Token} (TTFT) is primarily determined by the cumulative latency of these phases, significantly affecting the responsiveness of Serverless LLM services and user experience.

Prior studies have made efforts to reduce TTFT. SLLM~\cite{fu2024serverlessllm} improves \emph{Load} efficiency through CPU-based model caching and parallel parameter loading; 
Several works~\cite{criu,checkpointing,criugpu, serverless-coldstart-init-2} exploit Checkpoint/Restore In Userspace (CRIU) to bypass redundant context initialization in \emph{Init}; 
Medusa~\cite{medusa} uses offline materialization to pre-compute the information required for \emph{Profile}; 
Tidal~\cite{tidal} creates adaptive CUDA context templates to reduce the latency of kernel loading in \emph{Prefill}. Despite these notable advances, cold-start latency remains high, especially for large models.

To investigate the cold-start bottleneck of Serverless LLMs, we incrementally extended the widely used SLLM platform with state-of-the-art optimizations and evaluated models of varying sizes (ranging from OPT-1.3B to GPT-20B). 
As shown in Figure~\ref{fig:cold-start-optimizations}a, with CRIU-enabled checkpointing (\emph{SLLM-C}) and further with offline materialization (\emph{SLLM-CM}), the latencies of \emph{Init} and \emph{Profile} are significantly reduced. 
However, the latency of \emph{Load} remains high. 
Figure~\ref{fig:cold-start-optimizations}b shows that as model size increases, the latencies of other phases remain relatively stable, whereas the \emph{Load} latency scales almost linearly.
This indicates that the \emph{Load} latency is becoming the major performance bottleneck, which must be addressed. 

The key to reducing Load latency is to minimize the amount of data transferred, which can be achieved by retaining model parameters as much as possible in GPU memory.  
This idea is motivated by two observations: 
First, in Serverless LLM, model requests exhibit a strong locality. Our analysis on Serverless LLM workloads reveals a repeated loading pattern. Certain models are predictably loaded and evicted. 
By retaining such models completely or partially in GPU memory, the volume of data transfer across PCIe can be reduced, thereby reducing the data loading time.
Second, both model parameter sizes and KV cache requirements vary substantially across inference requests. In many cases, GPU memory is not fully used during a run. Exploiting this idle memory capacity opens an opportunity to leverage the unused GPU memory for retaining data.

Unfortunately, current Serverless LLM deployments operate under a restrictive assumption---once a model instance is scheduled to a GPU, it exclusively occupies the GPU memory for its entire service lifetime. 
During initialization, after the model parameters are loaded, most of the remaining GPU memory is reserved for the KV cache to reduce computation overhead. However, the actually required KV cache size is highly dependent on the input sequence length and inference batch size, which vary significantly across queries and over time. As a result, a large portion of the reserved GPU memory for KV cache remians unused.
When a model’s lifecycle ends, all GPU-resident data, including reusable model parameters, are discarded, even if subsequent requests require the same model. 
This over-conservative design, inherited from traditional single-model serving architectures where it works well, unfortunately results in severe resource underutilization and eliminates reuse opportunities in the Serverless LLM environment. To fully exploit GPU memory and reduce model-switching latency, {\it the deployment paradigm must evolve from an exclusive, single-model approach to a shared-memory, multi-model architecture}.

In this work, we present \textbf{Tangram}, a Serverless LLM framework that accelerates model loading through GPU memory reuse and affinity. 
Unlike the traditional exclusive resource-binding approach, Tangram enables parameters from multiple models to share GPU memory.
When loading a new model, it selectively evicts resident parameters based on size and access frequency. Frequently accessed (\emph{hot}) parameters are retained.
These preserved parameters are reused to serve subsequent requests for the same model, thereby reducing data transfer and model-switching latency.

To minimize model parameter loading overhead, Tangram employs a {\it tensor-level reuse strategy} along with a {\it unified memory pool} for memory allocation and reclamation.
For efficient memory management, 
Tangram formulates tensor reuse as the Multi-Choice Multi-Dimensional Knapsack Problem and applies a two-stage heuristic algorithm to guide allocation decisions.
Tangram also supports on-demand KV cache allocation by integrating the Unified Memory Pool with the ElasticKV engine and E-Attention kernel, enabling KV blocks to be allocated dynamically during runtime decoding.
To exploit GPU affinity, Tangram employs a {\it GPU-affinity–aware scheduler} that considers reusable parameters on GPUs when making scheduling decisions and selects the GPU with the highest expected loading efficiency.

We have implemented a prototype of Tangram based on SLLM~\cite{fu2024serverlessllm} and VLLM inference engine~\cite{vllm}. The implementation is open-sourced~\cite{tangramcode}.
We evaluate Tangram on a Serverless LLM platform using real-world workloads.
Experimental results show that Tangram accelerates model loading by up to $6.2\times$ and reduces the overall TTFT by $23\%\text{--}55\%$ compared to state-of-the-art approaches.


The remainder of this paper is organized as follows. Section~\ref{sec:background} introduces the background and motivation. Section~\ref{sec:Tangram} and ~\ref{sec:implementation} describe the design and implementation.
Section~\ref{sec:evaluation} presents the performance evaluation. Section~\ref{sec:related-work} reviews the related work. Section~\ref{sec:conclusion} concludes the paper.

\section{Background and Motivation}
\label{sec:background}

\subsection{Serverless LLM}
\label{sec:background-serverlessllm}

\begin{figure}
    \centering
    \includegraphics[width=0.5\textwidth]{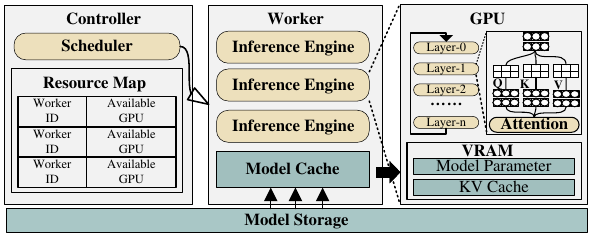}
    \caption{Serverless LLM Architecture.}
    \label{fig:serverlessllm}
\end{figure}

Serverless LLMs adopt a distributed architecture, where {\it central Controller} nodes manages the scheduling of model requests, and {\it Worker} nodes execute model inference, as illustrated in Figure~\ref{fig:serverlessllm}.
Controller nodes maintain a {\it resource map} recording the real-time status of worker nodes, including available GPUs and the execution state of each model instance.

A model must be registered before handling inference requests.
During {\it registration}, the user uploads the model parameters to the platform, which are then stored in persistent storage and assigned a unique {\it model ID}.
When an inference request for a model arrives, Controller nodes first check whether the model is already loaded in an inference engine. If so, the request is dispatched to that engine and queued for execution. Otherwise, the Controller invokes its scheduling algorithm to locate idle GPU resources. If no resources are available, the request must wait until the running tasks complete and GPU resources are released. 

When a model instance is allocated to a worker node, it runs exclusively on GPUs, depending on the size of the model. After completing its inference task, the instance enters a brief idle period to reduce cold-start frequency by keeping the model temporarily active. When the idle period ends, the instance is terminated and its parameters are evicted to release GPU resources.

\subsection{LLM Inference}
\label{sec:background-llm}

Once a model instance is scheduled to a worker node, it goes through a multi-phase initialization process to set up the inference environment and generate the first token. The initialization process mainly includes four phases:

\textbf{Init} establishes the runtime context for the target model, including Python library initialization, inference engine startup, and model architecture setup. Recent studies leverage CRIU checkpointing to preserve initialized contexts and restore them for subsequent deployments~\cite{checkpointing, criugpu}, thereby eliminating redundant overhead and reducing the latency of Init.

\textbf{Load} transfers model parameters to GPU memory, reconstructing them into tensors. SLLM~\cite{fu2024serverlessllm} optimizes this process by partitioning parameters into manageable chunks and caching them in CPU memory. These chunks are transferred to the GPU in parallel, improving Load efficiency by maximizing PCIe bandwidth utilization.

\textbf{Profile} conducts a preliminary inference run of the target model to determine the optimal memory allocation for various data components, such as the KV cache, activation data, and other intermediate tensors, which are required during model execution.
Medusa~\cite{medusa} improves this process by pre-computing model-specific profiles offline, thereby removing profiling overhead from the  critical path and further accelerating startup.

\textbf{Prefill} runs the entire input prompt in parallel to compute the KV pairs and generate the first token. Due to current GPU's lazy kernel loading approach, \emph{Prefill} typically includes kernel launch overhead. Tidal~\cite{tidal} proposes using an adaptive kernel template to reduce the launch overhead. 

After \emph{Prefill}, the model enters the \emph{Decode} stage, during which it generates tokens sequentially using all previously computed KV pairs. In each decoding step, the input passes through the stack of Transformer layers, where the attention mechanism computes intermediate representations until the final layer projects them into logits for the next token~\cite{atten-1, atten-2}. At the same time, KV pairs are also generated and stored for each new token.

During inference, a model maintains a KV cache to avoid recomputing KV pairs for previously generated tokens.  
However, storing this cache incurs substantial memory overhead, as its size depends on both {\it sequence length} and  {\it batch size}.
Sequence length is the total number of tokens generated during Prefill and Decode for a single inference request; the batch size is the number of inference requests processed simultaneously. Both factors vary dynamically, making it challenging to predict memory requirements.
To ensure stable performance, existing systems conservatively pre-allocate GPU memory for the maximum sequence length for each request.

\subsection{Performance Bottleneck and Opportunities}
\label{sec:background-opportunity}

To investigate cold-start bottlenecks, we have developed a prototype of Serverless LLM  that integrates three state-of-the-art optimizations: SLLM parallel loading~\cite{fu2024serverlessllm}, CRIU-based checkpointing~\cite{criu,checkpointing,criugpu}, and Medusa's offline profiling~\cite{medusa}, which we refer to as \emph{SLLM-CM}.

\begin{figure}[t]  
    \centering
    \begin{subfigure}[b]{0.49\linewidth}
        \centering
        \includegraphics[width=\linewidth]{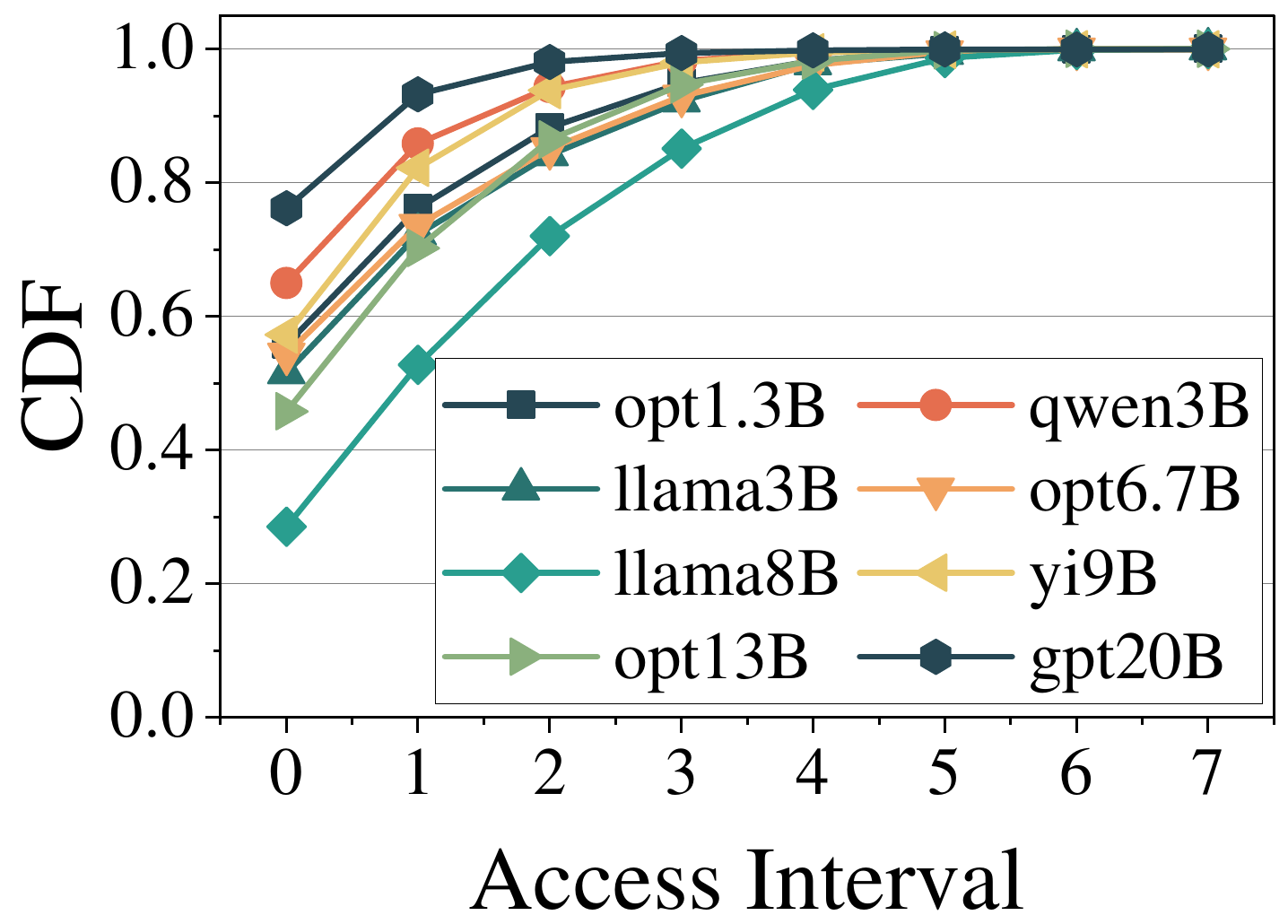}
        \caption{Model access interval.}
        \label{fig:moti-uniquemodel}
    \end{subfigure}
    \hfill
    \begin{subfigure}[b]{0.49\linewidth}
        \centering
        \includegraphics[width=\linewidth]{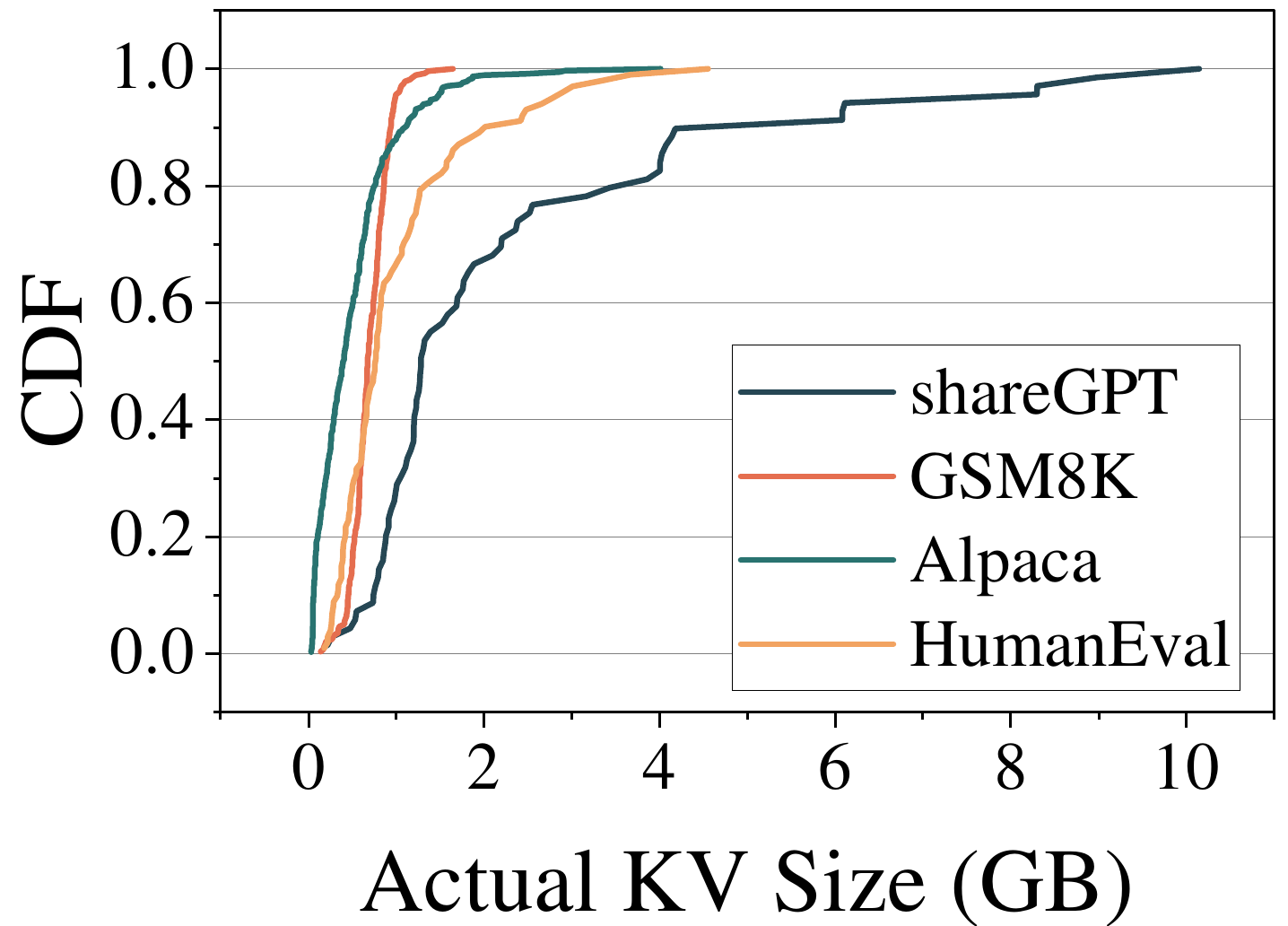}
        \caption{KV Cache size.}
        \label{fig:moti-memcons}
    \end{subfigure}
    \caption{(a) Model access intervals in Serverless LLM trace. (b) Measured KV cache size during inference (Batch Size=16).}
    \label{fig:motivation}
\end{figure}

We measure the TTFT breakdown of SLLM-CM across different model sizes. As shown in Figure~\ref{fig:cold-start-optimizations}, the integrated optimizations significantly reduce overall TTFT, but {\it Load} latency remains dominant and grows linearly with model size, revealing parameter loading as the primary bottleneck in cold starts. This paper explores opportunities for reducing Load latency through GPU memory reuse and GPU affinity, guided by two key observations:

\textbf{Observation~1: Model accesses exhibit locality.}
We find that in Serverless LLM platforms, the incoming inference tasks exhibit temporal locality. To investigate this, we use a real-world serverless trace~\cite{azura} to generate a sequence of access requests across eight different models. 
Figure~\ref{fig:moti-uniquemodel} illustrates the access interval of each model, where the interval is defined as the number of intervening models between two consecutive requests to the same model. 
The results show that most requests are consecutive, meaning that if the model parameters remain in GPU memory, subsequent requests for the same model could avoid redundant data transfers.

Traditional serverless deployments exploit this locality by configuring the \texttt{keep\_alive} parameter to retain active model instances for a specified period, e.g., 4 minutes~\cite{keep-alive-1,keep-alive-2, serverless-coldstart-cache}. 
However, the request interval for the same model is highly variable in real-world scenarios, making it challenging to select a static, pre-configured value that minimizes cold starts and prevents unnecessary GPU occupancy.

\textbf{Observation~2: Current KV cache allocation is overly conservative.}  
LLM inference engines typically pre-allocate the {\it maximum} GPU memory for the KV cache, without considering the dynamic variability in sequence lengths observed in real-world inference. 
As shown in Figure~\ref{fig:moti-memcons}, empirical measurements on the \texttt{Llama8B} model (batch size~16) across four representative datasets—ShareGPT~\cite{sharegpt}, GSM8K~\cite{gsm8k}, Alpaca~\cite{alpaca}, and HumanEval~\cite{humaneval}—reveal that actual KV cache consumption can vary by a factor of 25–137 across datasets.
While the KV cache may occupy nearly all available memory in worst-case scenarios, it usually consumes far less than the maximum allocation.


Based on the above observations, we have identified two important opportunities for optimizations: (1) Leveraging the temporal locality of inference tasks, retaining parameters in GPU memory even {\it after} the model instance is released can avoid the latency of reloading data in the next run. 
However, simply caching the last model’s parameters is insufficient, as access intervals vary significantly across models (Figure~\ref{fig:moti-uniquemodel}), particularly at interval~0.
If only retaining the last used model's parameters in GPU memory, 
models that have a relatively long reuse interval (e.g., \texttt{llama8B} in this example) would still be frequently loaded and evicted. 
(2) Allocating KV cache based on {\it actual demand} frees unnecessarily reserved GPU memory, which can then be used to retain model parameters.
This allows parameters of inactive model instances to be preserved in GPU memory.

\section{Design of Tangram}
\label{sec:Tangram}
In this section, we present the design details of Tangram. Our design aims to achieve three important goals.

\textbf{Aim \#1: Minimize model parameter loading.}
To accelerate model loading, it is crucial to eliminate redundant parameter transfers by exploring model temporal locality. Tangram seeks to minimize parameter loading by (i) selecting an appropriate reuse granularity, and (ii) identifying and utilizing the reusable parameters during model loading. 

\textbf{Aim \#2: Enable efficient memory management.}
To maximize reuse efficiency, Tangram aims to incorporate an efficient memory management mechanism that (i) prioritizes retaining more high-value model parameters, (ii) mitigates memory fragmentation caused by frequent model switching, and (iii) supports on-demand KV cache allocation.

\textbf{Aim \#3: Exploit GPU affinity.}
With memory reuse enabled, GPUs differ in their affinity for specific models based on the parameters already resident in memory. Tangram requires a GPU affinity–aware scheduler that assigns requests to selected GPUs to maximize the benefits of memory reuse and minimize end-to-end latency.

\begin{figure}[t]
    \centering
    \includegraphics[width=0.5\textwidth]{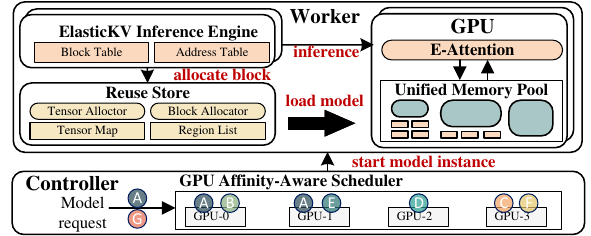}
    \caption{Tangram Architecture.}
    \label{fig:Tangram}
\end{figure}

Figure~\ref{fig:Tangram} illustrates the architecture of Tangram. 
Tangram employs a {\it Tensor-level Model Reuse and Loading} mechanism which retains the model tensors in reused GPU memory, and overrides the Load operation to only transfer the necessary tensors (Section~\ref{sec:Tangram-tensor}). 
To enable efficient memory management, Tangram integrates the {\it Reusable Memory Management}  (Section~\ref{sec:Tangram-memory}) to solve memory shortage problems at minimal cost within limited time, and leverages the {\it On-Demand KV cache Allocation} (Section~\ref{sec:Tangram-kvcache}) to further increase reusable GPU memory. Finally, Tangram also designs a {\it GPU Affinity-Aware Scheduler} (Section~\ref{sec:Tangram-affinity}) to allocate models to their favored GPUs. In the following sections, we introduce the details of each component.

\subsection{Tensor-level Model Reuse and Loading}
\label{sec:Tangram-tensor}

In traditional Serverless LLMs, model parameters are loaded and released at the {\it model} level, dictated by the exclusive use of the GPU by each model. 
However, adopting model as the basic unit leads to overly coarse-grained memory management, causing memory waste and poor utilization in Serverless LLM scenarios. 
A na\"ive alternative is to partition GPU memory into equally sized {\it pages} and map parameters to them, allowing page-level loading and eviction during model switches.
However, with a large number of concurrent hardware threads, GPU execution depends on coalesced accesses to global memory. 
Mapping parameters to non-contiguous physical pages breaks coalescing, wastes bandwidth, and could degrade performance by up to an order of magnitude~\cite{nvidianonesequence}. 
In addition, due to the overly small granularity, managing a large number of pages exacerbates the management cost.

We propose to realize parameter reuse at the {\it tensor} level. 
Tensors are the fundamental units for organizing and scheduling computation in LLMs. Reusing data in units of tensors can ensure that each tensor’s data remains contiguous, which helps preserve computational efficiency and avoid extensive modifications to existing GPU kernels. 
As a model typically comprises dozens of tensors, this approach is more fine-grained than model-level reuse, reaching a balance between fine-grained memory utilization and acceptable management complexity.

As illustrated in Figure~\ref{fig:Tangram}, Tangram deploys a \emph{Reuse Store} on each worker node to manage the reusable memory of each GPU (the Unified Memory Pool) and to support tensor-level model loading. For each GPU, the Reuse Store generates a unique fingerprint as the index for every tensor. A hash table–based \emph{Tensor Mapping} is maintained to keep track of each tensor's index (fingerprint) and its corresponding address in GPU memory.

During a model switch, once a model is assigned to a specific GPU, the Reuse Store first 
searches in the Tensor Mapping table for the model's tensors. 
For tensors not found, memory is allocated through the \emph{Tensor Allocator} in the Unified Memory Pool and the tensors are loaded from either the CPU-side Model Cache or the persistent Model Store. If the GPU does not have sufficient free memory space, the system evicts selected tensors from other ``inactive but resident'' models to free space. After all required tensors are loaded, the Reuse Store updates the tensor mapping and returns the actual GPU memory addresses of all tensors to the Inference Engine for computation.

\subsection{Reusable Memory Management}
\label{sec:Tangram-memory}

To enable tensor-level model reuse, 
the Reuse Store logically  divides the GPU memory space into a sequence of {\it regions}.
Each region is a contiguous memory space in the Unified Memory Pool; and the regions are chained together following their physical layout. 
A region is either {\it free} or {\it allocated}. 

During the model loading process, if a tensor is absent from the Tensor Map, the Tensor Allocator needs to select a free region to allocate memory for it. 
A commonly used strategy is the \emph{best-fit} approach, which assigns 
the smallest free region that can accommodate the tensor,
and if necessary, evicts existing tensors using an LRU policy. 
Each time when a tensor is allocated from a free region, a new, smaller free region could also be produced (for the remaining unused space). 
Many small, non-contiguous free regions are scattered in the memory space, none of which is large enough to hold a tensor’s contiguous memory requirements, even though the total free space is sufficient. In such cases, model loading could fail due to memory fragmentation.

The fragmented memory space must be reorganized, but this incurs non-trivial memory-copy overhead~\cite{defrag-1,defrag-2,defrag-3}. Tangram addresses this by modeling tensor reuse allocation as a Multi-Choice Multi-Dimensional Knapsack Problem (MCMDKP)~\cite{mcmdkp, mcmdkp-2} and applying a two-stage heuristic loading strategy to minimize overhead.


\subsubsection{Multi-Choise Multi-Demension Knapsack Problem.}
\label{sec:Tangram-reuse-knapsack}

When allocating space for new tensors, we can either (1) {\it evict} existing tensors to free regions or (2) {\it move} a tensor to merge adjacent free regions into a larger one. Eviction may cause additional latency if the evicted tensors are later accessed, while movement incurs immediate merge overhead, increasing loading latency. We formalize this process as a \emph{Multi-Choice Multi-Dimensional Knapsack Problem} (MCMDKP), formulated in Equation~\ref{eq:knapsack}.

\begin{equation}
    \label{eq:knapsack}
    \begin{aligned}
    \textbf{Given:}\;\;
    & T=\{t_1,\dots,t_N\},\; T'=\{t'_1,\dots,t'_M\},\; S, \\
    & s_i,\; s'_j,\; c_j,\; m_j. \\
    \textbf{Variables:}\;\;
    & x_i,y_j,z_j \in \{0,1\}. \\
    \textbf{Constraints:}\;\;
    & \sum_{i=1}^{N} s_i x_i \le S - \sum_{j=1}^{M} s'_j (1 - y_j), \\
    & y_j + z_j \le 1, \;\; \forall j, \\
    & \sum_{i=1}^{N} x_i = N. \\
    \textbf{Objective:}\;\;
    & \min \sum_{j=1}^{M} (c_j y_j + m_j z_j) \\
    \end{aligned}
\end{equation}

In the equation above, $T=\{t_1,\dots,t_N\}$ denotes the set of new tensors, where the total number of tensors is $N$, and each tensor $t_i$ of which needs to be allocated with space of size $s_i$; $T'=\{t'_1,\dots,t'_M\}$ denotes the set of existing tensors (total number is $M$) currently residing in GPU memory, each tensor $t'_j$ of which has size $s'_j$, eviction cost $c_j$, and merge cost $m_j$. 
The total size of the Unified Memory Pool is denoted by $S$.
We define three binary decision variables: $x_i=1$ if a new tensor $t_i$ is successfully allocated, $y_j=1$ if an existing tensor $t'_j$ is evicted, and $z_j=1$ if an existing tensor $t'_j$ is merged. 

The constraints in the equation guarantees that:
(1) the memory consumed by new tensors does not exceed the available GPU memory after counting in evicted tensors; 
(2) an existing tensor cannot be both evicted and merged simultaneously;
(3) all new tensors must be allocated with space. 
The objective is to minimize the overall allocation cost, defined as the weighted sum of eviction and merge costs across all existing tensors.

\subsubsection{Two-Stage Loading Heuristic.}
\label{sec:Tangram-reuse-algorithm}

The MCMDKP is NP-hard, as it can be reduced from the classical {\it 0--1 knapsack} problem. Tangram solves it with a two-stage heuristic algorithm that decomposes the multi-dimensional, multi-choice problem into stepwise subproblems.

\textbf{Stage~1: Minimal-Cost Eviction.}  
The goal of the first stage is to identify a set of tensors for eviction such that the total size of free regions is sufficient for the new tensors. 
Tangram defines the eviction cost $c_j$ for tensor $t'_j$ as:
{\small

\begin{equation}
    {c_j = p_m \cdot \left( \frac{s'_j}{b_m} \right) \cdot \alpha_m} ,
\end{equation}
}

\noindent where model $m$ is the model to which tensor $t'_j$ belongs.  
Here, $p_m$ denotes the tensor miss probability from model $m$, $s'_j$ the size of tensor $t'_j$, $b_m$ the model loading bandwidth, and $\alpha_m$ the model's loading-latency sensitivity.  
The value of $\alpha_m$ is specified by user during registration, since different models may have distinct responsiveness requirements~\cite{cold-start-sensitivity-1, cost-start-sensitivity-2}; It is set to 1 by default; for less latency-sensitive models, $\alpha_m$ can be set to a smaller value.

Tangram adopts a greedy strategy, which evicts tensors based on ascending eviction cost until the released size meets the requirement for loading the new tensors. In other words, we reclaim memory with the lowest-cost tensor. 

\begin{figure}[t]
    \centering
    \includegraphics[width=0.5\textwidth]{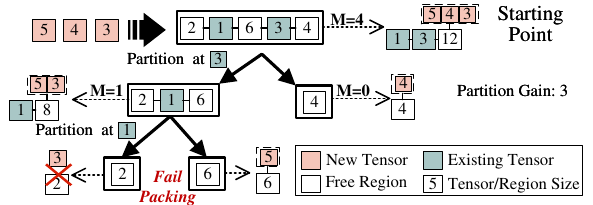}
    \caption{Partitioned-Gain Packing algorithm.}
    \label{fig:partitioned-bin-packing}
\end{figure}

\textbf{Stage~2: Partitioned-Gain Packing.}
Due to fragmentation, the total free space may be sufficient yet still unable to accommodate all new tensors.
A simple solution is to merge all free regions together to one end, forming a single range of contiguous free space, but this incurs significant merge cost.
Tangram implements a \emph{Partitioned-Gain Packing} algorithm to solve this problem~\cite{partitioned-bin-packing}. 
The objective is to fit new tensors directly in available free regions and perform merging {\it only} when necessary.

\begin{algorithm}[htb]
    \caption{Partitioned-Gain Packing Algorithm}
    \label{alg:recursive-split-packing}
    \KwIn{New tensors $T$ (sorted by size in descending order), Region List $R$}
    \KwOut{Allocation set $\mathcal{A}$ with subspaces and tensor subsets, and overall merge cost $\mathcal{M}$}
    
    \SetKwFunction{TryPacking}{TryPacking}
    
    $\mathcal{P} \leftarrow \{(R, T)\}$;
    $\mathcal{A} \leftarrow \emptyset$;
    $\mathcal{M} \leftarrow \text{size of all allocated regions in } R$\;
    
    \While{$\mathcal{P} \neq \emptyset$}{
        \ForEach{$(P, \mathcal{T})$ in $\mathcal{P}$}{
            $split\_occurred \leftarrow \textbf{false}$\;

            \ForEach{$t'_p \in P$}{
                Split $P$ at $t'_p$ into $(P_1, P_2)$\;
                $\mathcal{T}_1, \mathcal{T}_2 \leftarrow \TryPacking(\mathcal{T}, P_1, P_2)$\;

                \If{$(\mathcal{T}_1, \mathcal{T}_2) \neq (\emptyset, \emptyset)$}{    
                    $split\_occurred \leftarrow \textbf{true}$\;
                    Replace $(P, \mathcal{T})$ with $(P_1, \mathcal{T}_1)$ and $(P_2, \mathcal{T}_2)$ in $\mathcal{P}$\;
                    $\mathcal{M} \leftarrow \mathcal{M} - t'_p.size$;

                    \textbf{break}\;
                }
            }
            \If{$split\_occurred = \textbf{false}$}{
                Remove $(P, \mathcal{T})$ from $\mathcal{P}$\;
                Add $(P, \mathcal{T})$ to $\mathcal{A}$\;
            }
        }
    }
    
    \Return{$\mathcal{A}, \mathcal{M}$}\;
    
    \SetKwFunction{TryPacking}{TryPacking}
    \Fn{\TryPacking{$\mathcal{T}$, $P_1$, $P_2$}}{
    
    $\mathcal{T}_1, \mathcal{T}_2 \leftarrow \emptyset$\; 
    $C_1 \leftarrow P_1.capacity$; $C_2 \leftarrow P_2.capacity$\;

    \ForEach{tensor $t' \in \mathcal{T}$}{
        \If{$t'.size \geq \min(C_1, C_2)$}{ 
            \Return{$\emptyset, \emptyset$}\;  
        }
        
        \If{$C_1 \geq C_2$}{
            Add $t'$ to $\mathcal{T}_1$; $C_1 \leftarrow C_1 - t'.size$\;
        }
        \Else{
            Add $t'$ to $\mathcal{T}_2$; $C_2 \leftarrow C_2 - t'.size$\;
        }
    }
    
    \Return{$\mathcal{T}_1, \mathcal{T}_2$}\;
}

\end{algorithm}

The algorithm starts with the worst-case ``merge-all'' scenario, then explores feasible assignment schemes, evaluating their reductions in merge cost to identify the most cost-effective option. 
Thus, the problem of minimizing merge cost is reformulated as the problem of maximizing the assignment gain.
Since enumerating all allocation schemes is computationally intractable, we adopt a partitioning and bin-packing strategy to obtain a near-optimal solution.

{\bf Partitioning subspaces}. A \emph{subspace} ($P$) is defined as a consecutive set of regions that begins and ends with a free region, characterized by its \emph{capacity} $C$ (the total size of free and allocated regions) and \emph{maximum merge cost} $M$ (the total size of allocated regions). For example, in Figure~\ref{fig:partitioned-bin-packing}, the initial subspace has $M=4$ and $C=12$. A subspace can accommodate a set of new tensors ($\mathcal{T}$), if its capacity is at least their total size of $\mathcal{T}$, meaning that all tensors can be placed with no more than cost $M$. Our objective is to partition the subspace so that all new tensors are packed into free regions at the lowest possible merge cost.

Based on this idea, we employ Partitioned-Gain Packing (Algorithm~\ref{alg:recursive-split-packing}) to recursively search for an allocation plan with minimal merge cost. The algorithm maintains a pending set $\mathcal{P}$ of subspace–tensor pairs and an allocation set $\mathcal{A}$ of finalized allocations (line 1). For each pair $(P,\mathcal{T})$ in $\mathcal{P}$ (line 3), it traverses the allocated regions $t'_p$ in $P$ as potential \emph{partition points} (line 5), each splitting $P$ into $P_1$ and $P_2$ and yielding a \emph{partition gain} of $t'_p.size$ (line 7), since the tensor no longer contributes to the merge cost, thereby reducing $\mathcal{M}$ by $t'_p.size$ (line 11). At each candidate point, the \textsc{TryPacking} function attempts to divide $\mathcal{T}$ into $\mathcal{T}_1$ and $\mathcal{T}_2$ that fit into $P_1$ and $P_2$. Successful partitioning generates two new subspace–tensor pairs added to $\mathcal{P}$ for further recursion (line 10), while subspaces that cannot be partitioned are moved to $\mathcal{A}$ (line 12). The process terminates when $\mathcal{P}$ becomes empty.

{\bf Packing tensors in subspace}. Determining whether a partitioning attempt succeeds can be viewed as a bin-packing problem with two bins.
Tangram employs a variant of the \emph{Best Fit Decreasing} (BFD) policy~\cite{bin-packing-bfd}, which places the descending-sorted tensors into the subspace with the largest remaining space one by one (line-17 to line 27 in Algorithm~\ref{alg:recursive-split-packing}).  
If any tensor cannot fit into the subspace, the process terminates immediately and returns failure, meaning that the partitioning attempt at this partition point is unsuccessful.
Note that the tensors are only sorted once, and the packing process maintains this order across all generated subsets.
Given the small number of tensors, the sorting cost is negligible.


Figure~\ref{fig:partitioned-bin-packing} illustrates an example, in which the initial subspace has two candidate partition points with partition gains of 1 and 3. It first selects the point with gain~3; if succeeds, the subspace splits into two with tensors divided accordingly, and the total merge cost $M=4$ is reduced to~1. However, the left subspace fails to pack its subset, and the right subspace cannot be further partitioned. The algorithm therefore terminates with a final merge cost of~1.

\subsection{On-demand KV cache Allocation}
\label{sec:Tangram-kvcache}

\begin{figure}[t]
    \centering
    \includegraphics[width=0.5\textwidth]{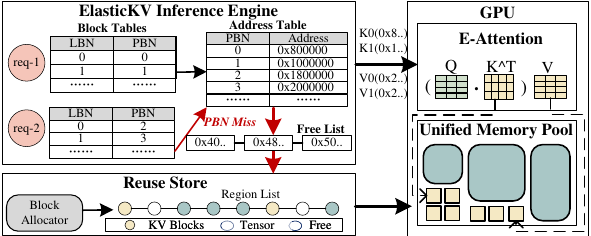}
    \caption{On-demand KV cache Allocation.}
    \label{fig:odkv}
\end{figure}


In traditional Serverless LLM architectures, once the model is loaded, all remaining GPU memory is reserved for the KV cache.
Although it guarantees inference performance in the worst-case scenario, such a conservative approach leads to significant memory waste due to workload variability, severely restricting the amount of memory available for Tangram to retain tensors for reuse.

Instead of pre-allocating KV cache space at initialization, Tangram utilizes the ElasticKV Inference Engine to allocate memory dynamically during inference. Since the KV size per token is fixed, the total cache requirement depends only on the current token count, which is determined by the input length in \emph{Prefill} and incremented by one token per step in \emph{Decode}. Before each phase, the ElasticKV Engine estimates the token count and allocates the exact required space.

In practice, multiple requests are often processed simultaneously~\cite{serverlessllm-batch, serverlessllm-batch-2, serverlessllm-batch-3}. 
Therefore, similar to conventional inference engines (e.g., vLLM~\cite{vllm}), the ElasticKV Inference Engine maintains a \emph{KV Block Table} for each request (as shown in the Figure~\ref{fig:odkv}). 
Each block stores the KV pairs of several consecutive tokens (e.g., 16). 
The Block Table records the mapping from a request’s {\it Logical Block Numbers} (LBNs) to the globally unique {\it Physical Block Numbers} (PBNs). 
Each PBN corresponds to a physical address in the Unified Memory Pool, which is tracked in the \emph{Address Table}. 
When a block becomes full, the ElasticKV Engine allocates a new physical block with the mapping updated in the Block Table.

The ElasticKV Inference Engine requests physical block space from the Unified Memory Pool by invoking the \emph{Block Allocator} in the Reuse Store, which allocates and reclaims blocks using the same Region List as the Tensor Allocator. 
However, in urgent situations with no free space, the Block Allocator directly reclaims tensors from inactive models based on the cost-aware eviction policy in Section~\ref{sec:Tangram-reuse-algorithm} 
to quickly reclaim memory and avoid impacting decode performance.
The Block Allocator marks regions assigned to KV blocks in the Region List and reclaims them collectively once the current model instance completes.


Before each inference begins, the ElasticKV Engine provides the Attention Kernel with all KV blocks of the current request along with their physical addresses for computation. 
Tangram implements an \emph{E-Attention Kernel} based on PagedAttention~\cite{vllm}, which executes attention at block granularity and retrieves KV pairs dynamically from the Unified Memory Pool using physical addresses provided by the ElasticKV Engine.


{\bf Optimizations}. On-demand KV cache allocation improves memory utilization but introduces runtime overhead.
To mitigate this, completed request blocks are retained in a \emph{Free List} rather than immediately returned to the Reuse Store, and new blocks are allocated from the Reuse Store only when the Free List is empty.
Since multiple batched inference tasks often run concurrently, the ElasticKV Engine merges their block-allocation requests before dispatch, amortizing overhead through batch operations.
In addition, using a larger block size further reduces allocation frequency; this granularity is defined at system initialization and applied consistently by ElasticKV and the Block Allocator for allocation and reclamation.

\subsection{GPU Affinity-Aware Scheduler}
\label{sec:Tangram-affinity}

In Serverless LLM, the {\it request scheduler} determines the GPU for each request, directly affecting cold-start latency. 
Conventional schedulers, e.g., SLLM~\cite{fu2024serverlessllm}, mainly consider CPU-side resources (Model Cache), leaving GPU-side resources overlooked. 
Tangram, however, retains inactive model tensors in GPU memory for reuse, so loading time depends on the amount of reusable tensors, making GPU memory state an essential factor in scheduling.

Tangram adopts {\it GPU affinity–aware scheduling} (Algorithm~\ref{alg:gpu-affinity-scheduling}), which receives a sequence of model requests (each with a model ID) and produces a mapping of models to scheduled GPUs. 
It first enumerates available GPUs (line~1) and then traverses requests, selecting the best GPU for each model (lines~2–12). 
For a given request, it first checks GPU feasibility (e.g., sufficient available memory, lines~5–6), measures the size of reusable tensors (line~7), and estimates the loading time (line~8). 
The GPU with the lowest expected latency is chosen (lines~9–10). 
If no idle GPU can accommodate the model, the request remains in the queue for future scheduling; otherwise, the model is paired with its best candidate GPU, which is subsequently removed from the available pool(lines~11–12). 
By prioritizing GPUs with a higher proportion of reusable tensors, Tangram effectively reduces loading time by exploiting memory locality.

\textbf{Loading Time Estimation.} Tangram estimates the expected loading time for a model-GPU pair, which serves as the foundation for scheduling decisions: 
\begin{equation}
    t_{load} = \frac{S - S'}{B}
\end{equation}
\noindent where $S$ is the model size, $S'$ is the size of reusable tensors on the GPU, and $B$ is the measured model-loading bandwidth.
The bandwidth depends on the model’s storage location: if it is in the Worker’s Model Store, the model must first be copied to the Model Cache before GPU loading.
Tangram leverages the asynchronous loading mechanism of SLLM~\cite{fu2024serverlessllm}, overlapping the transfer from the Model Store to the Model Cache with the transfer from the Model Cache to GPU memory.
Consequently, $t_{load}$ is determined only by the slower transfer medium.

\begin{algorithm}[t]
    \caption{GPU Affinity-Aware Scheduling}
    \label{alg:gpu-affinity-scheduling}
    \KwIn{$Requests$ \tcp*{list of model IDs}}
    \KwOut{$Schedules$ \tcp*{list of (model\_id, gpu)}}

    $GPUs \gets \text{AvailGPUs}()$;\quad $Schedules \gets [\,]$;
    
    \For{$m \in Requests$}{
      $best\_gpu \gets \varnothing$;\quad $best\_latency \gets +\infty$;
      
      \For{$g \in GPUs$}{
        \If{\text{CanRun}(m, g) = \text{False}}{\textbf{continue};}
        
        $reuse\_size \gets \text{GetReuseSize}(m, g)$;
        
        $lat \gets \text{EstimateLoadTime}(m, reuse\_size, g)$;
        
        \If{$lat < best\_latency$}{
            $best\_gpu \gets g$;\quad $best\_latency \gets lat$;
        }
      }
      
      \If{$best\_gpu \neq \varnothing$}{
        $Schedules.\text{push}(m, best\_gpu)$;
        $GPUs.\text{erase}(best\_gpu)$;
      }
    }
    
    \Return{$Schedules$};
\end{algorithm}

\section{Implementation}
\label{sec:implementation}

We implemented a fully functional Tangram prototype based on SLLM~\cite{fu2024serverlessllm} and VLLM~\cite{vllm} with 11{,}829 lines of CPU code and 725 lines of CUDA code, fully available as open source~\cite{tangramcode}. Specifically, Tangram allocates the Unified Memory Pool with \texttt{cuda\_malloc}, transfers tensors via \texttt{cuda\_mem\_copy}, and shares GPU memory handles with ElasticKV through the CUDA IPC API. To improve copy efficiency, adjacent tensors are grouped until exceeding an empirically chosen 8\,MB threshold. ElasticKV, built on VLLM, restores tensors with \texttt{Torch.from\_blob} and passes KV block addresses to E-Attention, extending PagedAttention with two CUDA kernels (\texttt{reshape\_and\_cache\_segment} and \texttt{segmented\_attention}) to enable physical-address–based KV cache access. Finally, the GPU Affinity-Aware Scheduler extends SLLM’s resource-aware scheduler by querying each worker’s Reuse Store for real-time reusable tensor sizes and schedules requests based on the estimated loading time.

\section{Evaluation}
\label{sec:evaluation}

\subsection{Experimental Setup}
\label{sec:evaluation-setup}

\textbf{Experimental Environment.}  
We evaluate Tangram’s performance with two setups. We use a single server equipped with NVIDIA L40 GPU (45\,GB VRAM), a 224-core Intel Xeon Platinum 8480+ CPU, 1\,TB DDR4 memory, and 7\,TB SSD storage for single-node studies. 
For scalability studies, we use a multi-GPU server with eight NVIDIA 4090 GPUs (24\,GB VRAM each) in a Ray-based~\cite{ray} distributed setup consisting of one controller node and eight worker nodes (one GPU per worker). The controller node has a 16-core Intel Xeon Platinum 8352V CPU and 16\,GB DDR4 memory, while each worker node has one NVIDIA 4090 GPU and 64\,GB DDR4 memory.
All experiments are conducted with Python~v3.10, CUDA~v12.3, PyTorch~v2.5.0, and Ray~v2.10.0.

\textbf{LLM Models.}  
Following the model size distribution reported by multi-model service platforms~\cite{hyperbolic} and prior work~\cite{yu2025prism}, we select popular models from the OPT~\cite{opt}, LLaMA~\cite{llama2}, Qwen~\cite{qwen2}, Yi~\cite{yi}, and GPT~\cite{gpt2} series. Among them, 30\% have 1--3B parameters, 60\% have 4--13B, and 10\% have 14--30B.

\textbf{Workloads and Datasets.}  
We adopt a similar approach as SLLM~\cite{fu2024serverlessllm} to generate workloads, combining the open-source serverless trace from Azura~\cite{azura, serverless-workload} with the model list to produce corresponding model functions~\cite{alpaserve}. 
Request arrivals follow a Gamma distribution. 
The original trace contains many consecutive model requests; to evaluate Tangram under varying locality conditions, we adjust the {\it Coefficient of Variation} (CV) of the Gamma distribution and reduce the proportion of consecutive requests when CV~$<1$ to create more challenging cases with lower locality. The rule is as follows: 
\textbf{L1-Locality}: CV~=~0.25, no consecutive requests;  
\textbf{L2-Locality}: CV~=~0.5, halved consecutive request length;  
\textbf{L3-Locality}: CV~=~1, original consecutive request;  
\textbf{L4-Locality}: CV~=~2, consecutive request.  
For each model request, we use real datasets as inputs:  
ShareGPT~\cite{sharegpt} (chat), GSM8K~\cite{gsm8k} (math), Alpaca~\cite{alpaca} (daily conversations), and HumanEval~\cite{humaneval} (programming).

\textbf{Comparison Baselines.}  
We compare Tangram with the latest Serverless LLM optimizations as follows. The baseline is \textbf{SLLM}~\cite{fu2024serverlessllm} that caches model parameters in CPU memory and improves loading efficiency via chunking and parallel transfer.  
\textbf{SLLM-C} extends SLLM with CRIU checkpointing~\cite{criu} to reduce the latency of Init. It creates and stores context-ready model images locally to resume initialization directly from these images.  
\textbf{SLLM-CM} is built on SLLM-C with further enhancement using Medusa's offline profiling~\cite{medusa} to eliminate the Profile latency.

To isolate GPU loading performance, we assume CPU memory is sufficient, so all missing tensors are served from CPU memory in SLLM variants and Tangram.

\subsection{Overall Performance}

\begin{figure*}[]
    \centering
    \includegraphics[width=1.0\textwidth]{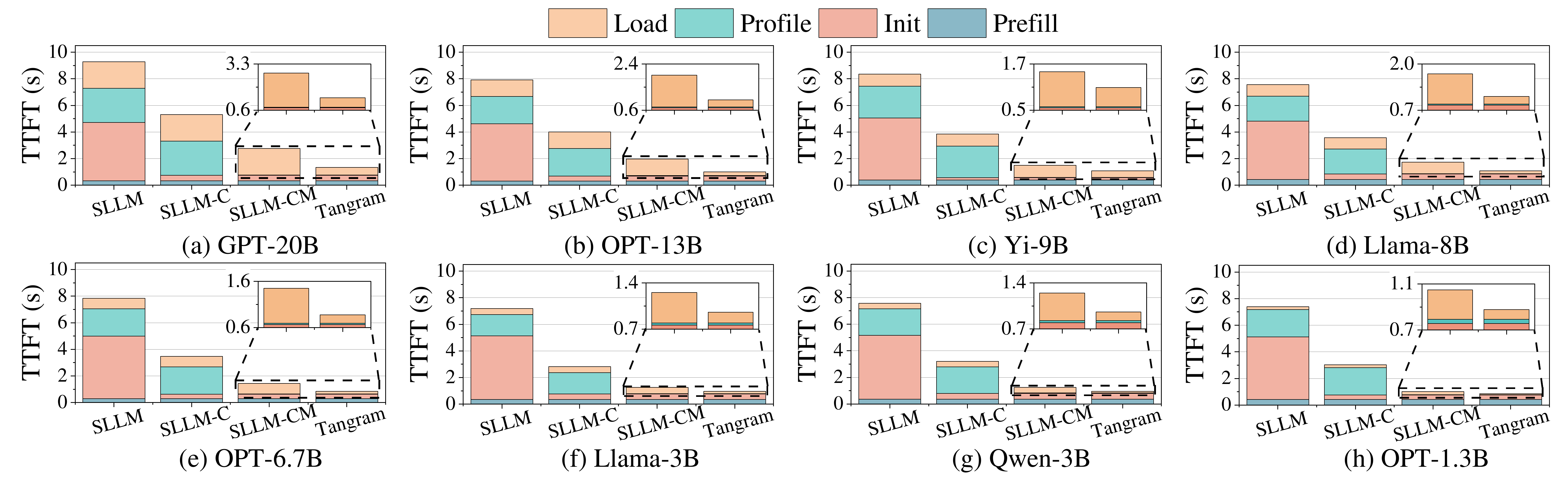}
    \caption{Overall performance: TTFT of different model under different Serverless LLM approaches.}
    \label{fig:overall-performance}
\end{figure*}

\begin{table}[]
    \centering
    \caption{Decode throughput (token/s) of different approaches.}
    \begin{tabular}{lcccc}
        \toprule
        \textbf{Model} & \textbf{SLLM} & \textbf{SLLM-C} & \textbf{SLLM-CM} & \textbf{Tangram} \\
        \midrule
        gpt20B   & 223.29  & 227.83  & 225.99  & 223.06 \\
        opt13B   & 303.03  & 299.54  & 297.33  & 293.22 \\
        yi9B     & 463.81  & 466.90  & 463.04  & 456.19 \\
        llama8B  & 509.02  & 499.07  & 507.82  & 497.84 \\
        opt6.7B  & 481.33  & 481.06  & 482.02  & 472.12 \\
        llama3B  & 620.24  & 618.30  & 620.40  & 611.80 \\
        qwen3B   & 552.30  & 554.35  & 551.43  & 538.11 \\
        opt1.3B  & 626.48  & 630.85  & 627.12  & 610.34 \\
        \bottomrule
    \end{tabular}
    \label{tab:overall-throughput}
\end{table}

We evaluate TTFT across different approaches and models. Figure~\ref{fig:overall-performance} shows the TTFT breakdown.  
With CRIU and offline profiling, SLLM-C and SLLM-CM significantly reduce the Init and Profile latency. 
As the model size increases, however, Load becomes the dominant bottleneck in SLLM-CM, accounting for up to 72\% of TTFT.  
Tangram mitigates this by reusing model parameters to avoid redundant data transfers, achieving a $1.8\times$--$6.2\times$ faster loading and reducing overall TTFT by 14\%--60\%.

Table~\ref{tab:overall-throughput} reports the decoding throughput of different approaches. SLLM, SLLM-C, and SLLM-CM achieve similar throughput since they share the same inference engine. Tangram’s on-demand KV cache allocation adds only minor overhead, with throughput loss under 3.2\%, which is negligible in practice.

\subsection{Performance Breakdown}

\begin{figure*}[ht]
    \centering
    \includegraphics[width=1.0\textwidth]{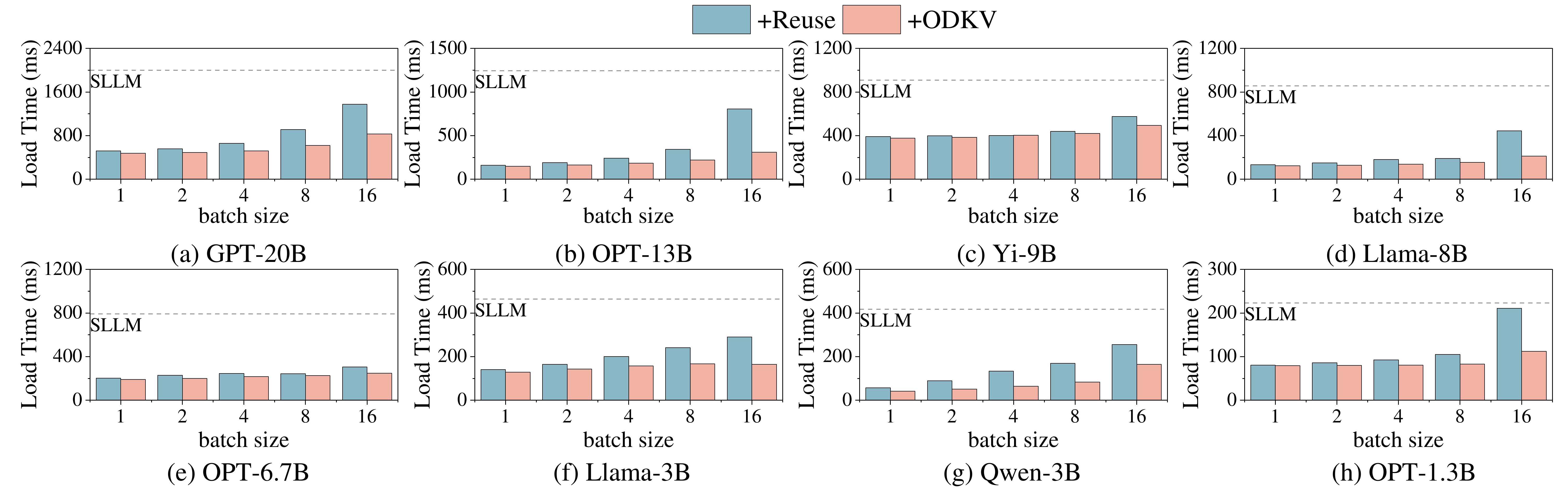}
    \caption{Performance breakdown: TTFT breakdown of Tangram under different batch size.}
    \label{fig:breakdown}
\end{figure*}

To analyze performance gains, Figure~\ref{fig:breakdown} shows Tangram's load performance breakdown across batch sizes. \textbf{“+Reuse”} denotes SLLM with Tangram’s Reuse Store. At batch size~1, “+Reuse” achieves a $2.3\times$--$7.6\times$ speedup across models. However, pre-allocating KV cache space reduces available GPU memory. In the worst case, the Memory Pool cannot fully accommodate the largest model, GPT-20B, forcing some tensors to be repeatedly loaded from CPU memory.

\textbf{“+ODKV”} augments “+Reuse” with Tangram’s on-demand KV cache mechanism, which allocates KV cache space from the Unified Memory Pool according to its actual needs, improving memory utilization.  
As batch size grows, “+ODKV” sustains strong performance, achieving $1.16\times$--$2.6\times$ speedups over “+Reuse” and $1.9\times$--$4\times$ over SLLM. These results highlight the importance of combining both optimizations to achieve the desired performance gains.

\subsection{Tangram Peformance Analysis}
\label{sec:Tangram-analysis}

\subsubsection{Tensor Allocation Policies}

\begin{figure}[t]
    \centering
    \includegraphics[width=0.5\textwidth]{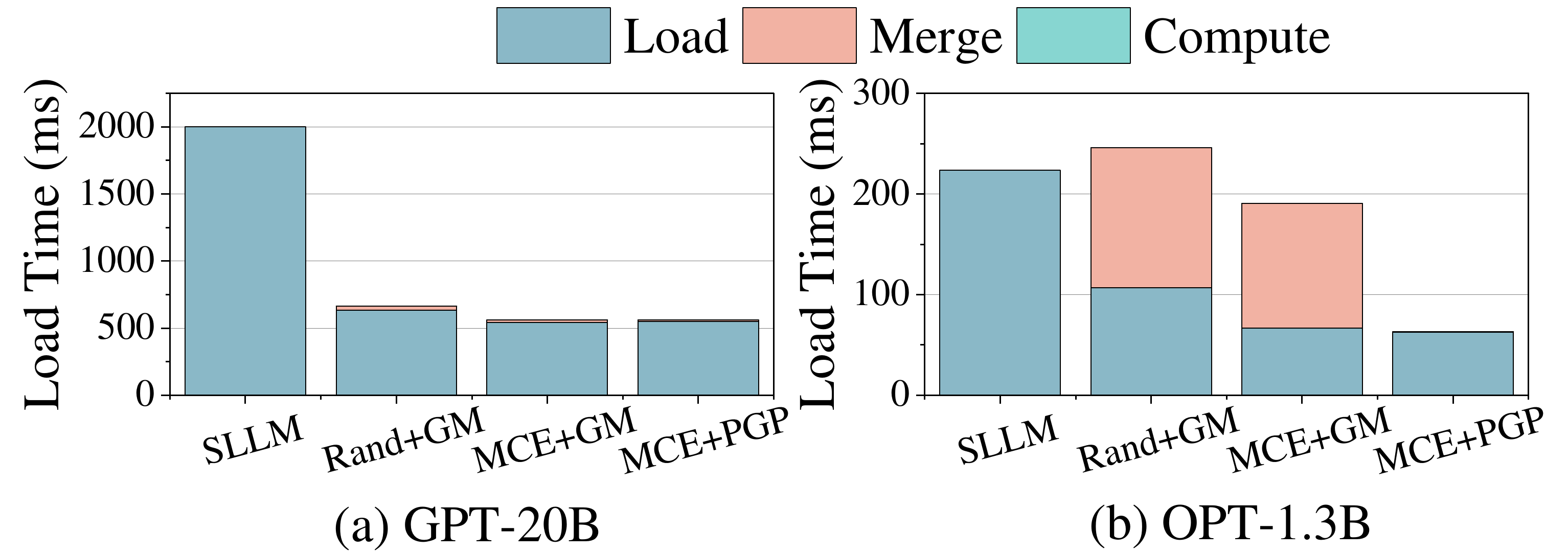}
    \caption{Effects of loading policies: Loading time breakdown for two representative model.}
    \label{fig:allocate-breakdown}
\end{figure}

We evaluate how different tensor allocation strategies affect Tangram’s loading performance.  
\textbf{“Rand+GM”} evicts tensors randomly when space is insufficient and merges free space by moving all the existing tensors (\emph{GlobalMerge}).  
\textbf{“MCE+GM”} applies Minimal-Cost Eviction in Section~\ref{sec:Tangram-reuse-algorithm} but still uses \emph{GlobalMerge}.  
\textbf{“MCE+PGP”} represents the Tangram strategy with Minimal-Cost Eviction and Partitioned-Greedy-Packing.  
For each strategy, we measure the latency overhead of \emph{Compute}, \emph{Merge}, and \emph{Load}.

Figure~\ref{fig:allocate-breakdown} presents results for two representative models: GPT-20B (large) and OPT-1.3B (small), with other models showing similar trends. Compared with SLLM, \emph{Rand+GM} reduces \emph{Load} time by up to 68\% through tensor reuse while Minimal-Cost Eviction further cuts \emph{Load} time by retaining more valuable tensors. However, in both \emph{GM} settings, merge overhead is nonnegligible, particularly for OPT-1.3B, as smaller models leave more existing tensors to be moved during Global Merge. This effect would even be amplified as Memory Pool becomes larger. Partitioned-Greedy-Packing alleviates this by minimizing unnecessary data movement, achieving a 93\% reduction in merge overhead. Across all three Tangram strategies, the additional \emph{Compute} latency remains under 1\,ms, making it negligible.

\subsubsection{On-demand KV cache Allocation}
\label{sec:Tangram-analysis-odkv}

\begin{figure}[t]
    \centering
    \begin{subfigure}[b]{0.49\linewidth}
        \centering
        \includegraphics[width=\linewidth]{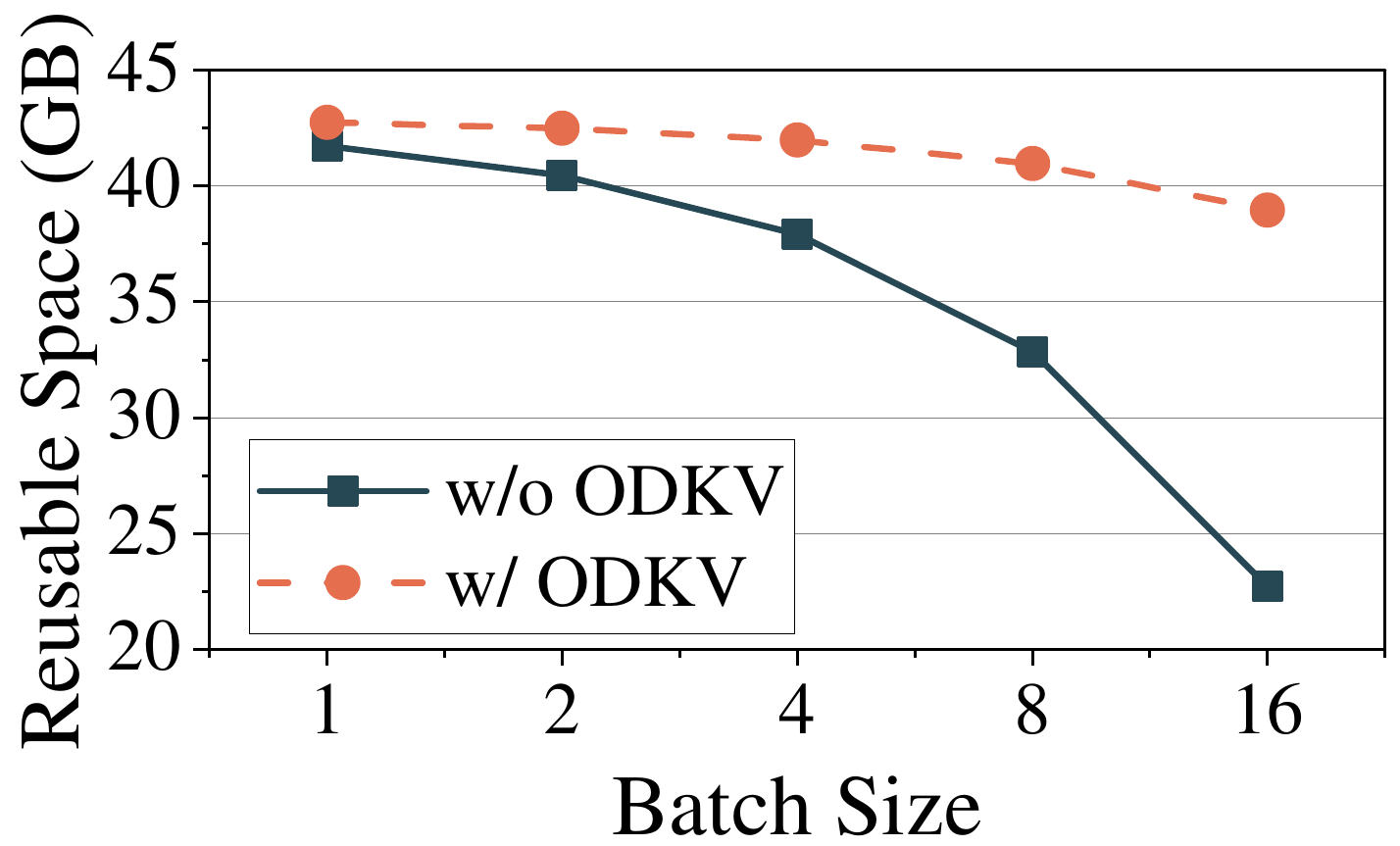}
        \caption{Tensor-Reusable Space}
        \label{fig:odkv-mem}
    \end{subfigure}
    \hfill
    \begin{subfigure}[b]{0.49\linewidth}
        \centering
        \includegraphics[width=\linewidth]{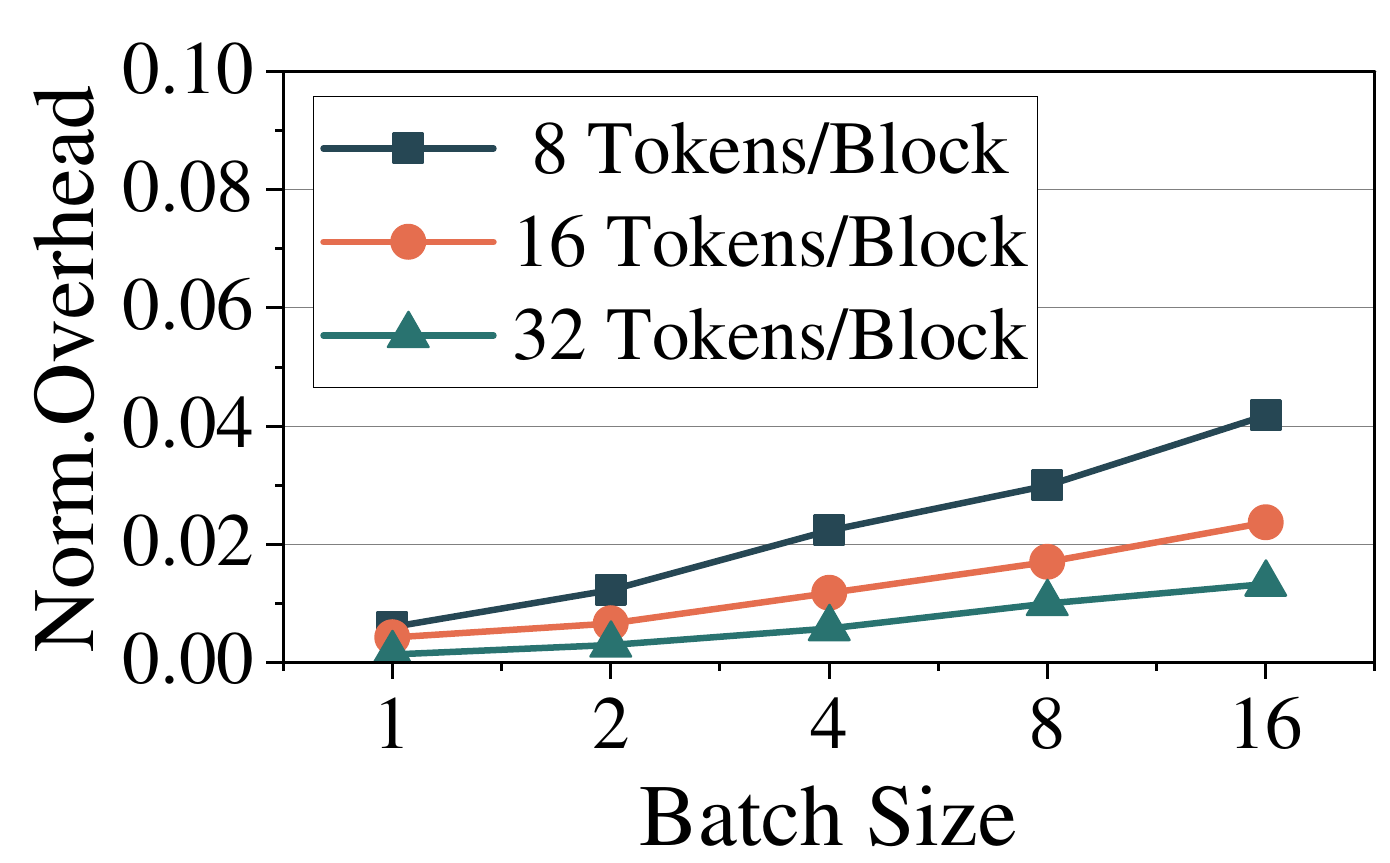}
        \caption{Overhead of ODKV}
        \label{fig:odkv-overhead}
    \end{subfigure}
    \caption{Effects of On-Demand KV cache Allocation: (a) Reusable space size under different batch sizes. (b) Overhead under different block sizes.}
    \label{fig:odkv-analysis}
\end{figure}

We evaluate the effect of On-Demand KV cache Allocation (ODKV) on tensor reuse space. Figure~\ref{fig:odkv-mem} shows how available space changes with batch size. In the {\bf ``w/o~ODKV''} setting, Tangram reserves KV cache space for each batch using the maximum sequence length.
In contrast, {\bf ``w/~ODKV''} allocates space based on actual sequence length, leaving more space for tensors.
As batch size grows, ODKV increases reusable space by 2.3\%--41\%.

Figure~\ref{fig:odkv-overhead} reports the overhead of ODKV, where the average allocation overhead is normalized by decode time.  
Tangram reduces real-time allocation costs through delayed release, a dedicated block allocator, and batched allocation.  
As a result, ODKV overhead remains below $0.041$ for a batch size of 16. Moreover, using coarser-grained block allocation further reduces overhead— increasing block size from 8 to 32 tokens lowers the overhead to $0.013$.

\subsection{Sensitivity Analysis}

\begin{figure}[t]
    \centering
    \includegraphics[width=0.5\textwidth]{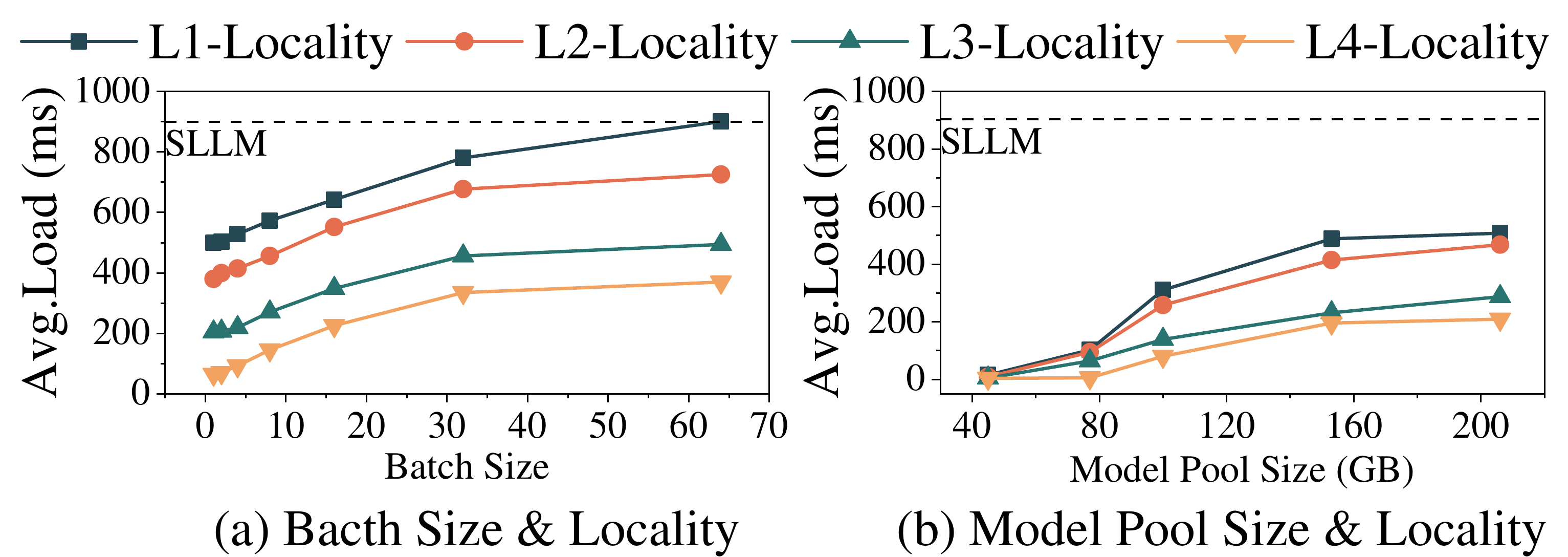}
    \caption{Sensitivity analysis: (a) Effects of workload locality and batch size (Model Pool Size = 120GB); (b) Effects of model pool size (Batch Size = 16).}

    \label{fig:sensitivity-analysis}
\end{figure}

We conduct a sensitivity analysis of Tangram’s performance. 
Figure~\ref{fig:sensitivity-analysis}a shows the impact of workload locality and batch size on loading performance.  
As locality decreases, Tangram benefits less from directly reusing parameters of the previously loaded model.  
Similarly, larger batch sizes force the Reuse Store to evict more tensors from the Memory Pool.  
At a batch size of 64, Tangram can rarely obtain reusable tensors, reaching its worst case when locality is L1 and batch size is 64, where its performance matches the SLLM baseline.  

Figure~\ref{fig:sensitivity-analysis}b shows the effect of model pool capacity on Tangram’s performance.  
In this experiment, we used a single GPU and gradually increased the total size of models scheduled to it.  
Tangram’s load latency initially increases but then stabilizes, and the higher locality is, the lower latency Tangram can achieve stably. 
Eventually, when the model pool exceeds 200\,GB, Tangram’s average load latency remains only 23\%--56\% of SLLM’s, as it maximizes Memory Pool utilization across varying request lengths to reuse tensors.

\subsection{Multi-GPU Performance}

\begin{figure}[t]
    \centering
    \includegraphics[width=0.5\textwidth]{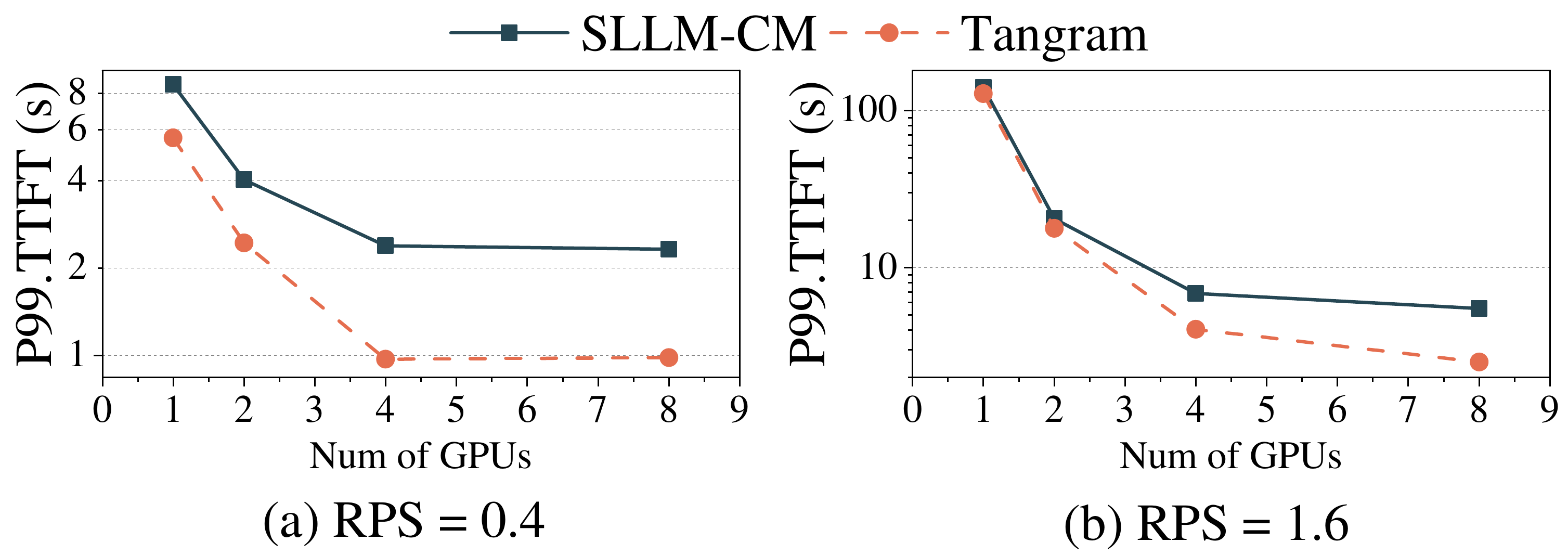}
    \caption{Multi-GPU Performance: 99th percentile tail latency of TTFT across different GPU configurations and request rates (Requests Per Second, RPS).}
    \label{fig:scalability-analysis}
\end{figure}

We evaluate the multi-GPU performance and scalability of Tangram under real workloads with various numbers of GPUs. 
Specifically, we control the request arrival rate, measured in requests per second (RPS), and scale the number of workers (each equipped with a single GPU) from 1 to 8. Figure~\ref{fig:scalability-analysis} reports the 99th-percentile latency of SLLM-CM and Tangram under two workload intensities.  

As shown in the figure, the tail latency of both SLLM-CM and Tangram decreases as the number of GPUs increases, with Tangram exhibiting a more pronounced reduction. At the low request arrival rate (RPS=0.4), the latency for both methods plateaus after the number of GPUs reaches 4, since additional GPUs provide no further benefit due to low intensity of incoming requests.
Nevertheless, under all GPU configurations, Tangram achieves noticeable latency reductions owing to its reuse mechanism, which directly lowers data transfer volume and thus shortens TTFT. 

At high request densities (RPS=1.6), limited GPU resources cause Tangram and SLLM-CM to converge in tail latency with more GPUs, as queueing delays dominate. With more GPUs provisioned, Tangram achieves 8\%–54\% lower latency by leveraging GPU affinity to select the optimal GPU for each model, compared with SLLM-CM’s random GPU selection.

\subsection{Overhead Analysis}
\label{sec:overhead-analysis}

We analyzed Tangram’s overhead on both Worker and Controller nodes. 
On Worker nodes, the primary storage cost arises from the CPU-side model cache, identical to SLLM, while the additional metadata overhead includes the \emph{Tensor Map} (52\,B per tensor) and the \emph{Region List} (48\,B per region), with typically fewer than $10^3$ entries in total. 
On Controller nodes, processing a model instance request entails querying Worker nodes for the status of real-time reusable tensors.
In our experiments with 8 Worker nodes, the RPC overhead of this query is 16.3\,ms, which is negligible compared to the second-level latency of model initialization.


\section{Other Related Work}
\label{sec:related-work}

\textbf{Optimization of Serverless LLM Loading.}  
SLLM~\cite{fu2024serverlessllm} leverages locality to cache model parameters in CPU memory and uses block-wise parallel transfer to improve PCIe bandwidth utilization.  
Prism~\cite{yu2025prism} coordinates multi-GPU transfers for further bandwidth gains, but both remain constrained by finite PCIe bandwidth versus growing model sizes.  
Sui et al.~\cite{sverlessllm-preload} and Tidal~\cite{tidal} aim to reduce loading latency through pre-loading, but they are limited by model size and stage-dependent constraints.
Tangram instead directly reduces the transfer volume during the Load phase via GPU memory reuse. 
Tangram is generally orthogonal to these techniques and can be integrated with them for additional performance gains, e.g., purposefully evicting certain Prefill tensors to create better overlapping opportunities for Tidal.

\textbf{GPU Memory Sharing.}  
VLLM~\cite{vllm} shares GPU memory across KV caches from different sequences but assumes a single model per system, reserving maximal KV space and wasting memory in multi-model Serverless LLM settings. 
Choi et al.~\cite{serverlessllm-share} first proposed spatially share a GPU to improve resource utilization for machine learning applications. 
Prism~\cite{yu2025prism} focuses on enabling multiple LLMs to run concurrently on a single GPU rather than optimizing cold-start latency. It manages memory at the model granularity, requiring full parameter loads for each launch and relying on CUDA for space management, which limits fragmentation control.
Tangram introduces tensor-level GPU memory reuse, a unified pool for model tensors and KV cache, and fragmentation-aware management, which together enable efficient memory utilization and reduced model loading latency in Serverless LLMs.

\section{Conclusion}
\label{sec:conclusion}

Tangram is a system designed to accelerate Serverless LLM loading through GPU memory reuse. 
It leverages unused memory to retain model parameters, unifies tensor and KV cache management in a shared memory pool, employs cost-aware allocation with on-demand KV allocation, and performs GPU-affinity–aware scheduling.
Our experiments demonstrate substantial improvements in reducing loading latency and TTFT, while scaling efficiently with multi-GPU resources at negligible overhead, 
which highlights GPU memory reuse as a promising direction for designing the next-generation Serverless LLM platforms.

\bibliographystyle{ACM-Reference-Format}
\bibliography{acmart}

\end{document}